\documentclass[a4paper,11pt]{article}
\pdfoutput=1
\usepackage{amsmath}
\usepackage{amssymb}
\usepackage{color}
\usepackage{cite}
\usepackage{hyperref}
\usepackage{graphicx}
 \usepackage{leftidx}
 \usepackage{subfig}

\makeatletter
\@addtoreset{equation}{section}
\renewcommand{\theequation}{\thesection.\@arabic\c@equation}
\makeatother
\makeatletter
\renewcommand\appendix{\par
  \setcounter{section}{0}%
  \setcounter{subsection}{0}%
  \gdef\thesection{Appendix \@Alph\c@section }
  \renewcommand{\theequation}
  {\Alph{section}.\arabic{equation}}
}
\makeatother

\def \be {\begin{equation}}
\def \ee {\end{equation}}
\def \ba {\begin{array}}
\def \ea {\end{array}}
\def \bea{\begin{eqnarray}}
\def \eea{\end{eqnarray}}

\def \a {\alpha}
\def \b {\beta}

\def \l {\lambda}

\def \s {\sigma}

\def \r {\rho}

\def \t {\tau}

\def \p {\partial}

\def \nn {\nonumber}

\def \hs {\hspace}

\def \inf {\infty}

\def \Tr {{\textrm{Tr}}}

\def \cL{\cal L}

\def \cW{\cal W}

\def \tr {{\textrm{tr}}}

\title{\textbf{Higher spin entanglement entropy at finite temperature with chemical potential}}
\author{
Bin Chen$^{1,2,3,4}$\footnote{bchen01@pku.edu.cn}\,
and
Jie-qiang Wu$^{1}$\footnote{jieqiangwu@pku.edu.cn}
}
\date{}

\begin{document}

\maketitle

\begin{center}
{\it
$^{1}$Department of Physics and State Key Laboratory of Nuclear Physics and Technology, Peking University, Beijing 100871, P.R.\! China\\
\vspace{2mm}
$^{2}$Collaborative Innovation Center of Quantum Matter, 5 Yiheyuan Rd, \\Beijing 100871, P.~R.~China\\
$^{3}$Center for High Energy Physics, Peking University, 5 Yiheyuan Rd, \\Beijing 100871, P.~R.~China\\
$^{4}$Beijing Center for Mathematics and Information Interdisciplinary Sciences, \\Beijing 100048, P.~R.~China
}
\vspace{10mm}
\end{center}

\begin{abstract}

It is generally believed that the semiclassical AdS$_3$ higher spin gravity could be described by a two dimensional conformal field theory with ${\cal{W}}$-algebra symmetry in the large central charge limit.  In this paper, we study the single interval entanglement entropy on the torus in the CFT with a ${\cW}_3$ deformation. More generally we develop the monodromy analysis to compute the two-point function of the light operators under a thermal density matrix with a ${\cW}_3$ chemical potential to the leading order. Holographically we compute the probe action of the Wilson line in the background of the spin-3 black hole with a chemical potential. We find exact agreement.

\end{abstract}

\baselineskip 18pt
\thispagestyle{empty}

\newpage

\section{Introduction}

The AdS/CFT correpondence provides a new tool to study the entanglement entropy.
 It was proposed by Ryu and Takayanagi that the entanglement entropy in the  conformal field theory(CFT) with a gravity dual can be evaluated by the area of a minimal surface in the bulk \cite{Ryu:2006bv}
 \be
 S_{HEE}=\frac{A}{4\pi G_N}.\label{RT}
 \ee
  The Ryu-Takayanagi(RT) formula (\ref{RT}) defines the holographic entanglement entropy(HEE), which implies a deep and intriguing relation between the entanglement and the quantum gravity. The HEE could be understood as a generalized gravitational entropy \cite{Lewkowycz:2013nqa}, as suggested by the similarity of (\ref{RT}) with the Bekenstein-Hawking entropy of the black hole.

  On the other hand, the holographic entanglement entropy opens a new window to study the AdS/CFT correspondence.  Especially in  the AdS$_3$/CFT$_2$ correspondence, the semiclassical AdS$_3$ gravity is dual to the large $c$ limit of the two-dimensional conformal field theory. In this context, under reasonable assumptions  the Ryu-Takayanagi formula has been derived in both the bulk \cite{Faulkner:2013yia} and the CFT \cite{Hartman:2013mia} in AdS$_3$/CFT$_2$. On the CFT side, the partition function of the $n$-sheeted Riemann surface could be simplified in the large $c$ limit. In fact, under this limit, the conformal block of multi-point functions could be dominated by the vacuum block, which allows one to solve the conformal block in the leading order using the monodromy techniques.  On the dual bulk side, loosely speaking, the classical handle-body solution ending on the $n$-sheeted Riemann surface could be constructed, and its on-shell action reproduces the leading order CFT partition function\cite{Krasnov:2000zq}. Moreover, it has been shown that the 1-loop correction to the RT formula in the bulk is captured exactly by the next-leading order contribution in the CFT partition function \cite{Headrick:2010zt,Barrella:2013wja,Chen:2013kpa,Chen:2013dxa,Perlmutter:2013paa,Chen:2014unl,Chen:2015uia,Chen:2015kua, Zhang:2015hoa}. This is due to the fact that the 1-loop partition function of any handle-body configuration \cite{Yin:2007gv,Giombi:2008vd} could be reproduced by the CFT partition function\cite{Chen:2015uga}.

  Furthermore the study of the entanglement entropy sheds light on the correspondence between the higher spin(HS) gravity and the CFT with ${\cal{W}}$ symmetry. In the first order formulation of the AdS$_3$ gravity, the theory could be rewritten in terms of the Chern-Simons(CS) theory with the gauge group $SL(2,R)\times SL(2,R)$\cite{Witten:1988hc}. By generalizing the gauge group from $SL(2,R)$ to $SL(N,R)$, the higher spin theory up to spin $N$ in AdS$_3$ could be constructed in the Chern-Simons formulation. The construction could be extended to the full higher spin algebra $hs[\l]$. More interestingly, by imposing the generalized Brown-Henneaux asymptotic boundary condition, the asymptotic symmetry group of the higher spin theory turns out to be generated by the $W_N$ algebra. This suggests that the higher spin AdS$_3$ gravity could be dual to a 2D CFT with ${\cal{W}}$-algebra\cite{Campoleoni:2010zq,Henneaux:2010xg}. One typical feature of the higher spin gravity is the loss of the diffeomorphism invariance. As a result, the usual geometrical notion like the horizon, the singularity and the area make no much sense. As a result, the RT formula (\ref{RT}) may not be able to compute the HEE in a higher spin theory.   More precisely, if one focus on the vacuum of the dual CFT, then the dual configuration is still gravitational and the higher spin fields appear only as the fluctuations around the classical configurations. In this case, one can still applies the RT formula and the higher spin fluctuations contribute only at the next-leading order\cite{Chen:2013dxa,Perlmutter:2013paa,Chen:2015uia}. However, if one considers the highly excited states with $W$ charge, then the dual configuration could be a higher spin black hole. Now the RT formula does not apply and one has to find a new way to compute  HEE.

  One promising proposal for HEE in the higher spin AdS$_3$ gravity is to use the Wilson line \cite{Ammon:2013hba,deBoer:2013vca}. As the theory is defined in the framework of Chern-Simons theory, it is natural to consider the Wilson line operator defined in terms of the gauge potential. By considering the Wilson line which ends on the branch points of the interval, it was proposed that the probe action of the Wilson line captures the entanglement entropy. As a consistency check, the entropy of the HS black hole has been reproduced.  Furthermore, it was shown in \cite{deBoer:2014sna} that the WL evaluated in a general asymptotically  AdS background captures correctly the correlation function in the dual CFT. This puts the WL proposal on a firmer footing.

  Let us review the work in \cite{deBoer:2014sna} in more details. On the dual field theory, the four-point correlation function, involving two heavy and two light operators, was considered. The heavy operator corresponds to the higher spin black hole or the conical defect with the higher spin hair. The light operator with conformal dimension $\Delta << c$ could be taken to be the twist operator in the $n \to 1$ limit. Therefore this four-point correlation function encodes the single-interval entanglement entropy of a highly excited states. From the operator product expansion (OPE), this four-point function can be decomposed into the contributions from the propagating states in different modules. The contribution from each module is called the conformal block. For a CFT with $\cW$ symmetry,  the states in the theory are classified by the representations of ${\cal{W}}$ algebra. Considering the large $c$ limit of the  CFT with  a sparse light spectrum,  only vacuum  conformal block dominates  the four-point function. By studying the monodromy problem of a differential equation,  the classical order of the conformal block can be computed.  Remarkably, as shown in \cite{deBoer:2014sna} explicitly this ${\cW}_3$ vacuum block can be computed even more efficiently by the bulk WL in the AdS$_3$ background corresponding to the heavy operator.

However, there are two subtleties in the study in \cite{deBoer:2014sna}.  First of all, the higher spin black hole solution usually contains two terms:
\be\label{holomorphic} a=a_z dz+a_{\bar{z}}d\bar{z}, \ee
for the holomorphic boundary condition\cite{Gutperle:2011kf}, and
\be\label{canonical} a=a_zdz+a_tdt, \ee
for the canonical boundary condition\cite{deBoer:2014fra}. The first term $a_z$ contains the charge, while the second term $a_{\bar{z}}$ or $a_t$ contains the chemical potential\cite{Henneaux:2013dra}. In the higher spin black hole solution, the asymptotic condition is different from the one for pure  AdS. The boundary condition corresponds to the higher spin deformation in the field theory. As shown in \cite{deBoer:2014fra}, there are two kinds of deformations to the CFT: the canonical deformation and the holomorphic deformation, corresponding to different  asymptotic boundary conditions. For the holographic entanglement entropy we should evaluate the probe action of the WL in terms of the gauge potential, whose  boundary condition should be in accord to the deformation in the field theory. However, in  \cite{deBoer:2014sna}, the chemical potential has been turned off in the holographic calculation.\footnote{This defect is not important for the entanglement entropy. When we evaluate the entanglement entropy in canonical deformation the two twist operator are at the same  time such that the chemical potential does not make effect. But for a general correlation function of two operators at different  time the chemical potential make a difference. We will go back to this problem later.} In other words, only  the gauge potential including only $a_z$ was discussed. Correspondingly,  there is no deformation in CFT side such that the monodromy analysis is easy to do. Generically speaking, when there is a deformation in CFT, the entanglement entropy is hard to compute \cite{Long:2014oxa,Datta:2014ska,Datta:2014uxa}. 

The second subtle point in \cite{deBoer:2014sna} is that the higher spin black hole microstate was regarded to be created by  a heavy operator in the CFT.  On the other hand, it is quite  often to use CFT at a finite temperature to represent a black hole. At the leading order, both pictures could be indistinguishable, but not at the quantum level \cite{Fitzpatrick:2016ive}. It would be interesting to study the HEE in a higher spin black hole background in the finite temperature picture. This is the issue we want to address in this paper.



In this paper, we study the single-interval entanglement entropy at a finite temperature and with a higher spin chemical potential.
  We use a thermal density matrix with a finite chemical potential to describe a higher spin black hole. There is moreover a canonical deformation term in the Hamiltonian,  corresponding  to the canonical boundary condition \cite{deBoer:2014fra} in the higher spin black hole solution.
 Our approach is different from the one in \cite{Long:2014oxa}. Instead of expanding the density matrix perturbatively in terms of the chemical potential, we treat the density matrix in a more exact way. This is feasible because we are only interested in the leading order result and we focus on the entanglement entropy rather than the R\'enyi entropy. Therefore we can use the saddle point approximation without worrying about the backreaction. In our case, the entanglement entropy is encoded in the two-point function of the twist operators under the density matrix. More generically we may consider the two-point function of two light primary operators with both conformal dimension and the spin-3 charge.  Instead of  studying the deformed theory directly, we take a picture-changing transformation and set the theory to a non-deformed theory. Under this picture transformation the two primary operators are transformed into two descendent operators, with the density matrix being invariant.  As the spatial  direction is compact, the correlation function is defined on a torus, we need to study the conformal block on the torus. We may insert a complete state bases at the thermal cycle. Basically the 2-point correlation on the torus reduces to  a sum of four-point functions which could be decomposed into the contributions from different propagating modules. We call the contribution from each module as a generalized conformal block. We can still use the monodromy analysis to study the leading order of generalized conformal block. Due to the presence of the chemical potential, we have one more differential equation, which correspond to evolving the operator by the higher spin charge. By solving the monodromy problem, we determine the leading-order correlation function of two general light operators under the thermal density matrix with the chemical potential.

 Furthermore, we discuss the HEE by computing the probe action of the Wilson line in the background of the higher spin black hole with the chemical potential in the canonical boundary condition. We find complete agreement with field theory correlator. The agreement between the 2-point function on torus and its holographic computation via Wilson line not only holds for the twist operators, but also for more general operators at different  time. On the bulk side, the picture-changing  transformation could be understood as the gauge transformation between different boundary conditions.




The remaining of the paper is organized as follows. In Section 2, we review the computation of the R\'enyi entropy at finite temperature. In particular, we give a derivation of the differential equation proposed in \cite{Barrella:2013wja} to study the conformal block on the torus. In Section 3, we study the two-point function of light operators on the torus with a chemical potential. We discuss the picture-changing transformation and introduce an auxiliary periodic coordinate with which the two-point function is defined with respect to a theory without deformation. We furthermore establish the differential equations of the wavefunction and the monodromy condition. In Section 4, we solve the monodromy problem to read the correlation function. In Section 5, we compute the correlation function holographically by using the Wilson line proposal. In Section 6, we ends with conclusion and discussion.

\section{R\'enyi entropy at finite temperature}

Entanglement entropy  measures the entanglement between the subsystem and its environment\cite{nielsen2010quantum}.
Assuming the whole system can be described by a density matrix $\rho$, we can define a reduced density matrix for the sub-system $A$ by tracing out the degrees of freedom in its environment $A^c$
\be \rho_A=\tr_{A^c} \rho.\ee
The entanglement entropy of subsystem $A$ is defined to be the Von Neumann entropy of the reduced density matrix
\be S_{EE}(A)=-\log \rho_A\log\rho_A. \ee
Moreover we can define the R\'enyi entanglement entropy
\be S_n(A)=-\frac{1}{n-1}\log \tr \rho_A^n, \ee
which allows us to read the entanglement entropy
\be\label{replica} S_{EE}=\lim_{n\rightarrow 1}S_n, \ee
if $n$ can be analytic extended to non-integer and the limit $n\rightarrow 1$ can be well taken.


The entanglement entropy and R\'enyi entropy can be computed by using the replica trick\cite{Holzhey:1994we}. The $n$-th R\'enyi entropy is given by
\be S_n=-\frac{1}{n-1}\log \tr \rho_A^n=-\frac{1}{n-1}\log \frac{Z_n}{Z_1^n}, \ee
where $Z_n$ is the partition function on an $n$-sheeted space-time connected with each other at the boundary of sub-region $A$. In the path integral formalism,  the partition function $Z_n$ can be taken in another way:  the field theory on the $n$-sheeted space-time is replaced by $n$-copies of the original theory on one-sheet spacetime with appropriate twisted boundary condition on the fields at the entangling surface. The entangling surface is at the boundary of sub-region $A$ at a fixed time, so is a surface of co-dimension 2. Circling around the entangling surface, the $i$-th copy of the field is connected with the $(i+1)$-th one.
Specifically, in two dimensional case, the entangling surface shrinks to some branch points, and the boundary condition on the fields at the branch points requires the introduction of the twist fields. The partition function can be computed by inserting the twist and anti-twist operators at the branch points in a orbifold CFT\footnote{The twist operator description can also be extended to higher dimensions, see \cite{Hung:2014npa}.}
\be\label{multi} \frac{Z_n}{Z_1^n}=\langle {\cal{T}}(z_1,\bar{z}_1){\cal{T}}(z_2,\bar{z}_2)...{\cal{T}}(z_{2N},\bar{z}_{2N})\rangle, \ee
where $N$ is the number of the intervals. In Eq. (\ref{multi}) , the correlation can be defined not only at the zero temperature but also at a finite temperature and even with a chemical potential as well.



In our case, we consider a two-point function on a torus, even with a higher spin current deformation. The correlation function can be decomposed into the generalised conformal block, whose leading order could be  computed by using the monodromy analysis. 
Before going into the details, we would like to give some general comments on the accuracy of the calculation. For a multi-point correlation function, we can use the operator product expansion (OPE) recursively and the correlation function can be decomposed into the contributions from different propagating  modules. If the correlation function is defined on a higher genus Riemann surface, we need to cut the Riemann surface open at some cycles and inserting the states in different modules.  We may just consider  one module  propagating in each OPE channel or one module at each cycle. This allows us to define  generalized conformal block, or conformal block on the torus. The multi-point function is a summation of the generalised conformal blocks. We have different ways to take the operator product expansion and cutting the cycles. For a theory with crossing symmetry and modular invariance, all of these expansions equal to each other, but with different convergent rates. It is believed that for a large $c$ theory, there is one kind of expansion in which the contribution from the vacuum module dominates and  the contribution from other modules is non-perturbatively suppressed in the large $c$ limit. Therefore even if we do not know the exact construction of the CFT dual to the AdS$_3$ gravity, we can still compute the correlation function reliably from the vacuum block as long as we find the proper channel. For different locations of the inserted operators and different Riemann surfaces, we may need to use different channels to expand the correlation function such that the vacuum module dominates in the expansion. There could be a phases transition at the parameter space when the expansion channel change. This effect is already known for Hawking-Page transition and in holographic entanglement entropy calculation \cite{Hartman:2013mia,Chen:2015kua}.

In the next subsections, we first review the conformal block of the four-point function on a complex plane as in  \cite{deBoer:2014sna}. Then we turn to the finite temperature case and show how to derive the partition function on a torus. Our discussion clarifies the proposal in \cite{Barrella:2013wja}.

\subsection{Conformal block from the monodromy}

Let us first consider the four-point function in full complex plane to show the general idea of the conformal block. For simplicity, we assume the symmetry of the theory is only generated by the Virasoro algebra. Our discussion follows \cite{Harlow:2011ny,Hartman:2013mia}. For the higher spin case, see Ref. \cite{deBoer:2014fra}.

For a general four-point function, we can evaluate it by inserting an identity  operator
\be \langle \phi_1(z_1)\phi_2(z_2)\phi_3(z_3)\phi_4(z_4)\rangle
=\sum_{\chi} \langle \phi_1(z_1)\phi_2(z_2)\mid \chi\rangle
\langle \chi \mid \phi_3(z_3)\phi_4(z_4) \rangle, \ee
where the states $\chi$ are normalized and orthogonal to each other. For a conformal field theory, the states can be classified by the representations of the conformal symmetry. The four-point function can be written as
\be \langle \phi_1(z_1)\phi_2(z_2)\phi_3(z_3)\phi_4(z_4)\rangle
=\sum_{\alpha}\sum_{\chi_{\alpha}}
\langle \phi_1(z_1)\phi_2(z_2)\mid \chi_{\alpha}\rangle
\langle \chi_{\alpha} \mid \phi_3(z_3)\phi_4(z_4) \rangle=\sum_{\alpha}{\cal{F}}_{\alpha}(x_i). \ee
The function ${\cal{F}}_{\alpha}(x_i)$ is called the conformal block, which is the conformal partial wave related to the representation $\alpha$.
The semi-classical limit is defined by taking $\Delta,c\rightarrow \inf$ with the ratio $\frac{\Delta}{c}$ being fixed. It is believed that under this limit the conformal block can be approximated to be
\be {\cal{F}}_{\alpha}(x_i)\approx e^{-\frac{c}{6}f(x_i)}, \ee
where $f(x_i)$ depends only on $\frac{\Delta}{c}$.

To determine the function $f(x_i)$, the standard way is to solve the monodromy problem. We first introduce a null state
\be
| \xi \rangle =(L_{-2}-\frac{3}{2(2\Delta_{\psi}+1)}L_{-1}^2) \mid \hat{\psi}\rangle, \label{nullstate}\ee
where
\be \Delta_{\psi}=\frac{1}{16}(5-c+\sqrt{(c-1)(c-25)}). \ee
In the large $c$ limit, $\Delta_{\psi}\rightarrow -\frac{1}{2}-\frac{9}{2c}$ and the null states goes to
\be\label{null1} | \xi \rangle=(L_{-2}+\frac{c}{6}L_{-1}^2\mid \hat{\psi} \rangle. \ee
Inserting the null state into the correlation funciton and defining
\be\label{psi} \psi(z)=\frac{\sum_{\chi_{\alpha}}\langle \phi_1(z_1)\phi_2(z_2)\mid \chi_{\alpha}\rangle
\langle \chi_{\alpha}\mid \hat{\psi}(z)\phi_3(z_3)\phi_4(z_4)\rangle}
{\sum_{\chi_{\alpha}}\langle \phi_1(z_1)\phi_2(z_2)\mid \chi_{\alpha}\rangle
\langle \chi_{\alpha} \mid \phi_3(z_3)\phi_4(z_4)\rangle}, \ee
\be\label{T} T(z)=\frac{\sum_{\chi_{\alpha}}\langle \phi_1(z_1)\phi_2(z_2)\mid \chi_{\alpha}\rangle
\langle \chi_{\alpha}\mid \hat{T}(z)\phi_3(z_3)\phi_4(z_4)\rangle}
{\sum_{\chi_{\alpha}}\langle \phi_1(z_1)\phi_2(z_2)\mid \chi_{\alpha}\rangle
\langle \chi_{\alpha} \mid \phi_3(z_3)\phi_4(z_4)\rangle}, \ee
we find that the decoupling of the null state leads to
\be\label{diff} \psi^{''}(z)+\frac{6}{c}T(z)\psi(z)=0, \ee
in the large $c$ limit, where by using the Ward identity, the stress tensor is of the form
\be\label{wald} T(z)=\sum_{i} \frac{h_i}{(z-z_i)^2}+\frac{1}{z-z_i}\frac{\partial}{\partial z_i}\log {\cal{F}}. \ee
In Eq. (\ref{psi}),  there is  a term
$\langle \chi_{\alpha}\mid \hat{\psi}(z)\phi_3(z_3)\phi_4(z_4)\rangle$. Because that  $\hat{\psi}$ is a null state, it leads to a differential equation. Solving the differential equation, we can get a monodromy when $z$ moves around $z_3$ and $z_4$. As the monodromy is the same for all of the $\chi_{\alpha}$, this indicates that $\psi(z)$ in (\ref{psi}) has such a monodromy as well.  With this monodromy condition and (\ref{diff}), we can solve all the coefficients in the stress tensor (\ref{wald}) and  fix the conformal block up to a constant. For the entanglement entropy, we need to consider vacuum conformal block. In this case the monodromy around $z_3$ and $z_4$ is trivial.

\subsection{Conformal block at finite temperature}

In this subsection, we discuss the  conformal block at a finite temperature. The thermal density matrix is
\be
\r_{thermal}=e^{2\pi i\t {\cL}_0-2\pi i{\bar \t} {\bar {\cL}}_0},
\ee
where
\bea
{\cL}_0&=&-\frac{1}{2\pi}\int_0^{2\pi} \hat T(w)dw, \notag \\
\bar{\cL}_0&=&-\frac{1}{2\pi}\int_0^{2\pi}{\hat {\tilde{T}}}(\bar{w}) d\bar{w}.
\eea
Consider a multi-correlation function of the primary fields $\phi_j$ with the conformal dimension $h_j$ on a torus. The torus is characterized by the moduli $\t$, and  is doubly periodic
\be
z \sim z+2\pi, \hs{3ex} z \sim z+2\pi \t .
\ee
By inserting a complete set of states  we change the correlation function on the torus to  a summation of the correlator on a cylinder
\bea \langle \prod_j \phi_j(z_j) \rangle \mid_{\tau}
&=&\Tr \left(e^{2\pi i\t {\cL}_0-2\pi i{\bar \t} {\bar {\cL}}_0}\prod_j \phi_j(z_j)\right)\nn\\
&=&\sum_{h,k,\bar{k}}\langle h,k,\bar{k}\mid \prod_j \phi_j(z_j) \mid h,k,\bar{k}\rangle
e^{2\pi i\tau(h+k-\frac{c}{24})-2\pi i \bar{\tau}(\bar{h}+\bar{k}-\frac{c}{24})}, \eea
where the index $h$ denotes the different primary modules propagating on the torus and $k$ and $\bar{k}$ denote the descendants in that module. Under the conformal transformation the different modules do not mix with each other so we define a finite conformal block which only sum over the states in one module
\bea {\cal{F}}(\tau,h;z_j,h_j;h_{p,r})
=\sum_{k}\langle h,k \mid \prod_j \phi_j(z_j) \mid h,k \rangle_{h_{p,r}}
e^{2\pi i\tau (h+k-\frac{c}{24})}, \eea
where we only consider holomorphic part.
It is a multi-point conformal block on the cylinder, with two descendants operators at the  past infinity and the future infinity respectively, and $h_{p,r}$ denote the conformal dimension of the propagators.

Let us focus on the holomorphic sector and derive  the Ward identity on the conformal block following the paper \cite{Eguchi:1986sb}. Consider the correlation function
\bea \sum_k \langle h,k \mid \hat T(z)\prod_j \phi_j(z_j) \mid h,k \rangle_{h_{p,r}} e^{2\pi i\tau (h+k-\frac{c}{24})}. \eea
Even though we only sum over one module on the torus,  the above function should be periodic in both direction. The periodic condition $z\rightarrow z+2\pi$ is trivial. For $z\rightarrow z+2\pi\tau$, we have
\bea\label{periodicity} \hat T(z) \sum_{k} \mid h,k \rangle \langle h,k \mid e^{2\pi i\tau (h+k-\frac{c}{24})}
= \sum_{k} \mid h,k \rangle \langle h,k \mid T(z+2\pi\tau) e^{2\pi i \tau(h+k-\frac{c}{24})} .\eea
 Therefore we find that
\bea &~&\sum_{k} \langle h,k \mid \hat T(z) \prod_j \phi_j (z_j) \mid h,k \rangle_{h_{p,r}}
e^{2\pi i\tau (h+k-\frac{c}{24})} \nn \\
&=&\sum_j \sum_n \frac{h_j}{4\sin^2 \frac{1}{2}(z-z_j+2\pi n\tau)}
\sum_{k} \langle h,k \prod_j \phi_j (z_j) \mid h,k \rangle_{h_{p,r}}
e^{2\pi i\tau (h+k-\frac{c}{24})} \\
&~&+\sum_j \sum_n \frac{1}{2} \cot \frac{1}{2}(z-z_j+2\pi n\tau) \frac{\partial}{\partial z_j}
\sum_{k} \langle h,k \prod_j \phi_j (z_j) \mid h,k \rangle_{h_{p,r}}
e^{2\pi i\tau (h+k-\frac{c}{24})}
+f(\tau), \nn\eea
where
\bea \sum_n \frac{1}{4\sin^2 \frac{1}{2}(z-z_j+2n\pi\tau)}&=&\frac{1}{4\pi^2}\wp (\frac{z-z_j}{2\pi}\mid \tau)+\frac{1}{2\pi^2}\eta_1(\tau), \notag \\
 \sum_n \frac{1}{2} \cot \frac{1}{2}(z-z_j+2n\pi\tau) &=&\frac{1}{2\pi}\zeta(z-z_j\mid\tau)-\frac{1}{2\pi^2}\eta_1(\tau)(z-z_j). \eea
To fix the $f(\tau)$, we need to take an integral along the spatial  direction. Considering that
\be \int_0^{2\pi} dz \hat T(z)=-4\pi^2 L_0+\frac{c}{6}\pi^2. \ee
we get
\be f(\tau)=2\pi i\frac{\partial }{\partial \tau} {\cal F}(\t). \ee

Similarly, by inserting the null state $|\xi\rangle$ in the correlation function as in (\ref{nullstate}),  we have a differential equation
\bea &~&\left\{-\frac{3}{2(2\Delta+1)}\frac{\partial}{\partial z^2} +2\eta_1(\tau) \Delta +\sum_j\left(\frac{1}{4\pi^2}\wp(\frac{z-z_j}{2\pi}\mid \tau)+\frac{1}{2\pi^2}\eta_1(\tau)\right)h_j
\right. \nn\\
&~&+\sum_j\left(\frac{1}{2\pi}\zeta(\frac{z-z_j}{2\pi}\mid \tau)-\frac{1}{2\pi^2}\eta_1(\tau)(z-z_j)\right) \frac{\partial}{\partial z_j}
 \nn\\
&~&\left. +2\pi i\frac{\partial}{\partial \tau}\right\}
\sum_k \langle h,k \mid \psi(z) \prod_j \phi_j(z_j) \mid h,k \rangle e^{2\pi i \tau(h+k-\frac{c}{24})}=0,
\eea
where $\psi(z)$ is the vertex operator for the state $\mid \hat{\psi}\rangle$.
Defining the function
\be \Psi\equiv \frac{\sum_k\langle h,k \mid \psi(z)\prod_j\phi_j(z_j) (z_j) \mid h,k \rangle e^{2\pi i \tau(h+k-\frac{c}{24})}}
{\sum_k\langle h,k \mid \prod_j\phi_j(z_j) (z_j) \mid h,k \rangle e^{2\pi i \tau(h+k-\frac{c}{24})}} \ee
which is assumed to be order $c^0$, and taking the large $c$ limit, we have the equation
\bea\label{equa}
& & -\frac{c}{6}\frac{\partial}{\partial z^2}\Psi
+\sum_j\left((\frac{1}{4\pi^2}\wp(\frac{z-z_j}{2\pi}\mid\tau)+\frac{1}{2\pi^2}\eta_1(\tau))h_j
+(\frac{1}{2\pi}\zeta(\frac{z-z_j}{2\pi}\mid\tau)-\frac{1}{2\pi^2}\eta_1(\tau)(z-z_j))\frac{\partial}{\partial z_j} \log {\cal{F}}\right)\Psi\nn\\
& & \hs{7ex}
+2\pi i\frac{\partial}{\partial \tau} \log {\cal{F}}\Psi=0 \eea
This is the equation (14) given in \cite{Barrella:2013wja}. Here we give a field theory derivation for that equation.

We need to fix the monodromy condition around the propagator with
\bea
M=\lim_{c\rightarrow \inf}\left(
\begin{array}{ccc}
e^{\pi i(1+(1-\frac{24h_{p,r}}{c})^{\frac{1}{2}})} &0 \\
0 & e^{\pi i(1-(1-\frac{24h_{p,r}}{c})^{\frac{1}{2}})}
\end{array}
\right),
\eea
and the monodromy around the spatial  cycle
\bea
M=\lim_{c\rightarrow \inf}\left(
\begin{array}{ccc}
e^{\pi i(1-\frac{24h}{c})^{\frac{1}{2}}} &0 \\
0 & e^{-\pi i(1-\frac{24h}{c})^{\frac{1}{2}}}
\end{array}
\right).
\eea
The matrix denote the transformation for two independent solutions of the equation (\ref{equa}) up to conjugate,  when the argument moves around the two cycles.

\section{Correlation function at finite temperature with higher spin deformation}

We now turn to  compute the entanglement entropy at a finite temperature with a finite chemical deformation. The entanglement entropy of a single interval could be read from  the two-point function of  two primary twist operators  in this system. The spatial direction of the torus is $-L/2 \leq \s \leq L/2$.  We define
\bea {\cal{L}}_0&=&-\frac{1}{2\pi} \int_{-\frac{L}{2}}^{\frac{L}{2}} \hat T(z)d\s, \notag \\
 {\cal{W}}_0&=&\frac{1}{2\pi} \int_{-\frac{L}{2}}^{\frac{L}{2}} \hat W(z)d\s, \notag \\
 \bar{\cal{L}}_0&=&-\frac{1}{2\pi} \int_{-\frac{L}{2}}^{\frac{L}{2}} {\hat {\bar{T}}}(\bar{z})d\s, \notag \\
 \bar{\cal{W}}_0&=&\frac{1}{2\pi} \int_{-\frac{L}{2}}^{\frac{L}{2}} {\hat {\bar{W}}}(\bar{z})d\s,   \eea
where the integral is over the real axis $-L/2 \leq \s \leq L/2$, and the Hamiltonian for the non-deformed theory is
\bea\label{H0P0} H_0&=&{\cal{L}}_0+\bar{\cal{L}}_0, \notag \\
 P_0&=&{\cal{L}}_0-\bar{\cal{L}}_0. \eea
For a theory with a higher spin current deformation,  the modified Hamiltonian is
\be\label{modify} H=H_0-\frac{2\pi i\alpha}{\beta}{\cal{W}}_0+\frac{2\pi i\bar{\alpha}}{\beta}\bar{\cal{W}}_0. \ee
In a deformed theory the system is evolved by this Hamiltonian.

\subsection{Picture-changing transformation}
In the modified system, we can define the Euclidian version of the two-point function at a finite temperature with a non-zero potential $\Phi$ conjugate to the momentum
\be\label{twopoint0} \frac{1}{Z}\Tr \left(\rho \phi_1^r(t_{E,1},\sigma_1) \phi_2^r(t_{E,2},\sigma_2)\right), \ee
where
\bea\label{content} \rho&=&e^{-\beta H+i\Phi P}, \notag \\
\phi^r(t_E,\sigma)&=&e^{t_E H}\phi^r(0,\sigma)e^{-t_E H}
=e^{t_E H}e^{-i\sigma P_0}\phi^r(0,0)e^{i\sigma P_0}e^{-t_E H}. \eea
The superscript $r$ denote that the operator is evolved by the modified Hamiltonian. In the finite temperature system, the operator is doubly periodic
\be\label{pe} \phi^{r}(t_E+\beta,\sigma+\Phi)=\phi^{r}(t_E,\sigma+L)=\phi^{r}(t_E,\sigma). \ee
For the discussion we can also define an operator evolved by the original Hamiltonian without deformation, as
 \be\label{phi}
 \phi(z,\bar{z})=e^{-iz{\cal{L}}_0}e^{i\bar{z}\bar{\cal{L}}_0}\phi(0,0)e^{iz{\cal{L}}_0}e^{-i\bar{z}{\cal{L}}_0}, \ee
where we have introduced the complex coordinate
\bea &~&z=\sigma+it_E, ~~~\bar{z}=\sigma-it_E. \eea
 If we regard the ${\cal{W}}_0$ $\bar{\cal{W}}_0$ term in Eq. (\ref{modify}) as an interaction, then the operator $\phi^r$ in (\ref{content}) is the operator in the Hamiltonian picture and the operator  $\phi$ in (\ref{phi}) is the operator in the interaction picture.
We can recombine the chemical potential $\Phi$ and the inverse temperature as a parameter in the complex coordinate
\bea
&~&2\pi \tau=\Phi+i\beta,~~~2\pi\bar{\tau}=\Phi-i\beta. \eea
Note that here we do not normalize the spatial direction so that the complex quantity $\t$ is not the moduli of the torus.

From  (\ref{H0P0}) and (\ref{modify}), we can rewrite the density matrix as
\be\label{density} \rho=e^{2\pi i\tau{\cal{L}}_0+2\pi i\alpha{\cal{W}}_0}
e^{-2\pi i\bar{\tau}\bar{\cal{L}}_0-2\pi i\bar{\alpha}\bar{\cal{W}}_0}.\ee
The  operators in the Hamiltonian picture and interaction picture are related to each other as
\bea \phi^r(\tau,\sigma)&=&e^{-\frac{2\pi i\alpha}{\beta}t_E{\cal{W}}_0}
e^{\frac{2\pi i\alpha}{\beta}t_E\bar{\cal{W}}_0}
e^{-iz{\cal{L}}_0}e^{i\bar{z}\bar{\cal{L}}_0}\phi^r(0,0)
e^{iz{\cal{L}}_0}e^{-i\bar{z}\bar{\cal{L}}_0}
e^{\frac{2\pi i\alpha}{\beta}t_E{\cal{W}}_0}
e^{-\frac{2\pi i\alpha}{\beta}t_E\bar{\cal{W}}_0}  \notag \\
&=&e^{-\frac{2\pi i\alpha}{\beta}t_E{\cal{W}}_0}
e^{\frac{2\pi i\alpha}{\beta}t_E\bar{\cal{W}}_0}
\phi(z,\bar{z})
e^{\frac{2\pi i\alpha}{\beta}t_E{\cal{W}}_0}
e^{-\frac{2\pi i\alpha}{\beta}t_E\bar{\cal{W}}_0} , \eea
where we have used the relation
\bea\label{evolveL} [i{\cal{L}}_0,\phi(z,\bar{z})]&=&-\frac{\partial}{\partial z}\phi(z,\bar{z}), \notag \\
{[}i\bar{{\cal{L}}}_0,\phi(z,\bar{z}){]}&=&\frac{\partial}{\partial \bar{z}}\phi(z,\bar{z}). \eea
The relation (\ref{evolveL}) can be proved as follows.
By the path integral we have
\bea\label{com} [i{\cal{L}}_0,\phi(z)]&=&[-\frac{i}{2\pi}\int_{-\frac{L}{2}}^{\frac{L}{2}} \hat T(z^{'})dz^{'},\phi(z)]
=\langle \frac{i}{2\pi}\oint dz^{'} \hat T(z^{'})\phi(z,\bar{z})\rangle \notag \\
&=& \frac{i}{2\pi}\oint dz^{'}\sum_{m=-\inf}^{\inf}\frac{\hat L_{m}}{(z^{'}-z)^{m+2}}\phi(z) \notag \\
&=&-(\hat L_{-1}\phi)(z)=-\partial \phi(z).
\eea
Thus, we see that  the translation along $z$ is  induced by the conserved charge ${\cal{L}}_0$. Similarly we may consider the evolution with respect to the charge ${\cal{W}}_0$ as well
\bea
{[}i{\cal{W}}_0,\phi(z){]}&=&[\frac{i}{2\pi}\int_{-\frac{L}{2}}^{\frac{L}{2}} \hat W(z^{'})dz^{'},\phi(z)]
=\langle -\frac{i}{2\pi}\oint dz^{'} \hat W(z^{'})\phi(z,\bar{z}) \rangle \notag \\
&=& -\frac{i}{2\pi}\oint dz^{'}\sum_{m=-\inf}^{\inf}\frac{\hat W_{m}}{(z^{'}-z)^{m+3}}\phi(z) \notag \\
&=&(\hat W_{-2}\phi)(z). \eea
By  introducing two other auxiliary coordinates $y, \bar y$, we can define
\be\label{new} \phi(z,y;\bar{z},\bar{y})\equiv
e^{-i{\cal{W}}_0 y}e^{i{\bar{\cal{W}}}_0\bar{y}}\phi(z,\bar{z}) e^{i{\cal{W}}_0y}e^{i\bar{\cal{W}}_0\bar{y}}, \ee
then we have
\be\label{ph}
\phi^r(t_E,\sigma)=\phi(z,y;\bar{z},\bar{y})
\ee
with
\be y=\frac{2\pi \alpha}{\beta}t_{E}~~~\bar{y}=\frac{2\pi \bar{\alpha}}{\beta}t_E. \ee
To compute the single-interval entanglement entropy,  $\phi_1$ and $\phi_2$ are taken to be the twist and anti-twist operators at the branch points respectively. Both operators are primary. With (\ref{ph}), we can write (\ref{twopoint0}) as
\be\label{twopoint1} \Tr \left(\rho \phi(z_1,y_1;\bar{z}_1,\bar{y}_1)\phi(z_2,y_2;\bar{z}_2,\bar{y}_2)\right).\ee
The operator $\phi(z_i,y_i;\bar{z}_i,\bar{y}_i)$ can be regarded to be a descendant operator inserted at $(z_i,\bar{z}_i)$, which is evolved by the non-deformed Hamiltonian ${\cal H}_0$.

Up to now, we have transformed the correlation function of two primary operators at a finite temperature in a deformed theory to the correlation function of  two descendant operators in a non-deformed theory under a density matrix at a finite temperature and with a finite chemical potential. We can regard this transformation to be a picture-changing transformation in quantum theory.  In Eq. (\ref{modify}),
$-\frac{2\pi i\alpha}{\beta}{\cal{W}}_0+\frac{2\pi i\bar{\alpha}}{\beta}\bar{W}_0$ can be regarded as an interaction term. In Eq. (\ref{new}) $\phi(z_i,y_i;\bar{z}_i,\bar{y}_i)$ on the left hand side can be regarded to be in the Heisenberg picture and its evolution is respect to the Hamiltonian with interaction,  and the operator $\phi(z,\bar z)$ on the the right hand side can be regarded to be in the interaction picture. In the Heisenberg picture we compute the correlation function of two primary operators in the deformed theory, while in the interaction picture we compute the correlation function of two descendant operators in non-deformed theory.  In  Section 5, we will show that the picture transformation here corresponds to the gauge transformation in the bulk theory. Different pictures here correspond to different boundary conditions in the bulk solutions.

With the relations (\ref{evolveL}) and (\ref{new}) it is easy to prove
\be e^{-i{\cal{L}}_0z_1}e^{-i{\cal{W}}_0y_1}\phi(z,y)e^{i{\cal{L}}_0z_1}e^{i{\cal{W}}_0y_1}=\phi(z+z_1,y+y_1). \ee
Furthermore using (\ref{pe}) or (\ref{density}) and (\ref{twopoint1}), we have
\be \langle \phi(z+2\pi i\tau,y+2\pi i \alpha)...\rangle|_{\tau,\alpha}=
\langle \phi(z,y)...\rangle|_{\tau,\alpha}, \label{generalperiodic}\ee
which is a generalized version of cyclic boundary condition in the thermal direction.  However one should be aware that this periodicity is only true for a complete theory. That means we need to sum over all  contributions from  different channels with proper combination between the holomorphic and anti-holomorphic part. If we consider only one conformal block, the periodicity may break down. For example in \cite{Faulkner:2013yia,Barrella:2013wja} the second order differential equation has been defined for the wavefunction of a multi-point function as Eq. (\ref{diff}) and Eq. (\ref{equa}). However, it was shown that the solution is not single-valued along the non-trivial cycle. Furthermore because the conformal block only contain holomorphic part, even in trivial cycle, the conformal block may have an extra phase, as shown in \cite{Barrella:2013wja}. 

\subsection{Monodromy problem}

In this subsection, we will show how to expand the correlation function (\ref{twopoint1}) in terms of  the generalized conformal block and set up the monondromy condition to compute  the generalized conformal block from propagating vacuum module states. In the semi-classical limit, we assume that the  propagating vacuum module dominates the contribution. In a theory with ${\cal W}_3$ symmetry, the vacuum module include the excitations of Virasoro generators and $W_3$ generators acting on the vacuum. 

\begin{figure}[tbp]
  \centering
  \subfloat[Numerator]{\includegraphics[width=6cm]{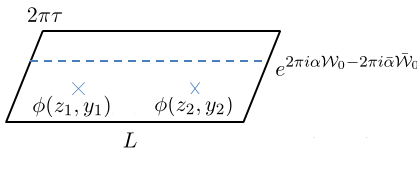}
  \label{numerator}}
   \subfloat[Denominator] {\includegraphics[width=4cm]{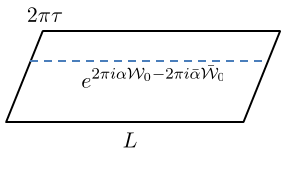}
  \label{denominator}}\\
  \caption{The  correlation functions with a non-local operator inserting at the dashed line. The operator can be moved to another time slice. The local operators are set below the nonlocal operator. }\label{correlation1}
  \end{figure}

By the path integral the function (\ref{twopoint1}) can be normalized to be
\bea\label{twopointT} C_2&=&\frac{\Tr e^{2\pi i\tau {\cal{L}}_0+2\pi i\alpha {\cal{W}}_0
-2\pi i\bar{\tau} \bar{\cal{L}}_0-2\pi i\bar{\alpha}\bar{\cal{W}}_0}
\phi(z_1,y_1;\bar{z}_1,\bar{y}_1)\phi(z_2,y_2;\bar{z}_2,\bar{y}_2)}
{\Tr e^{2\pi i\tau {\cal{L}}_0+2\pi i\alpha {\cal{W}}_0
-2\pi i\bar{\tau} \bar{\cal{L}}_0-2\pi i\bar{\alpha}\bar{\cal{W}}_0} } \notag \\
&=&\frac{\langle e^{2\pi i\alpha{\cal{W}}_0-2\pi i\bar{\alpha}\bar{\cal{W}}_0}
\phi(z_1,y_1;\bar{z}_1,\bar{y}_1)\phi(z_2,y_2;\bar{z}_2,\bar{y}_2)\rangle_{2\pi\tau,2\pi\bar{\tau}}}
{\langle e^{2\pi i\alpha{\cal{W}}_0-2\pi i\bar{\alpha}\bar{\cal{W}}_0}\rangle_{2\pi\tau,2\pi\bar{\tau}}}.
\eea
The correlation functions in the second line are defined on a  torus with $(L,2\pi\tau)$ being its periods. The $e^{2\pi i\alpha{\cal{W}}_0-2\pi i\bar{\alpha}\bar{\cal{W}}_0}$ is a non-local operator inserting at a time slice.
In this correlation function, the local operator can be continuously deformed in any contour away from the locations of the other operators, and the expectation value is continuously changed along this contour. When the contour crosses a non-local operator, the situation becomes subtle. In the case at hand, the non-local operators induce a jump of the operators. More precisely, we have the relation
\bea\label{cross} &~&e^{2\pi iz_1{\cal{L}}_0}{\phi}(z,y)={\phi}(z-z_1,y)e^{2\pi iy_1{\cal{L}}_0}, \notag \\
&~& e^{2\pi iy_1{\cal{W}}_0}{\phi}(z,y)={\phi}(z,y-y_1)e^{2\pi iy_1{\cal{W}}_0}. \eea
This means that crossing an operator $e^{2\pi iz_1{\cal{L}}_0}$ has the effect of  evolving $-z_1$ along the $z$ direction and crossing an operator $e^{2\pi iy_1{\cal{W}}_0}$ has the effect of  evolving $-y_1$ in the $y$ direction. These equation can be written in the path integral formalism as
\bea\label{evolvepath} &~&\langle e^{2\pi iz_1{\cal{L}}_0}{\phi}_{\mbox{\tiny lower}}(z,y)\cdots\rangle
=\langle{\phi}_{\mbox{\tiny upper}}(z-z_1,y)e^{2\pi iy_1{\cal{L}}_0} \cdots \rangle, \notag \\
&~& \langle e^{2\pi iy_1{\cal{W}}_0}{\phi}_{\mbox{\tiny lower}}(z,y) \cdots \rangle
=\langle {\phi}_{\mbox{\tiny upper}}(z,y-y_1)e^{2\pi iy_1{\cal{W}}_0} \cdots\rangle, \eea
where the subscript "lower" or "upper" denotes the operator $\phi$ is below or above the non-local operators.  Because the operator ${\cal{W}}_0$ commutes with the Hamiltonian, the inserted non-local operator $\alpha {\cal{W}}_0$  can be moved to any imaginary time slice if the movement don't touch other operators. Therefore, the correlation functions in the numerator and the denominator in Eq. (\ref{twopointT}) are represented respectively as in (\ref{numerator}) and (\ref{denominator}) in Fig.\ref{correlation1}.




As in previous section, both the numerator and denominator of (\ref{twopointT}) are the correlation functions on the torus. By inserting a complete basis in thermal cycle, they can be decomposed into contributions from the states in different modules. Furthermore for the two operators in the numerator we can  take an OPE and the expansion can be decomposed into the contribution from different modules. For each choice of modules in the OPE and in thermal cycle, the contribution defines a generalised conformal block. The numerator and denominator can be written as a summation of generalised conformal blocks from different modules.  In the large central charge limit, we assume that  the generalised conformal block from the  vacuum module  dominates the  contribution and the ones with other modules are  non-perturbatively suppressed. We note that the jump from the crossing a non-local operator (\ref{cross} ,\ref{evolvepath}) remains in the conformal block.


To determine this conformal block, we may use the monodromy analysis as before. However, due to the presence of ${\cal{W}}_3$, there is one more differential equation to consider. Introduce a primary state
$\mid \hat \psi \rangle$ such that
\bea L_0 \mid \hat \psi\rangle&=&-\mid \psi \rangle, \notag \\
W_0 \mid \hat \psi \rangle &=&\frac{1}{3} \mid \psi \rangle. \eea
In its descendants there are null states at  level 1, 2, 3. In the large $c$ limit, they are
\bea\label{nullvector} &~&(L_{-1}+2W_{-1}) \mid \hat \psi \rangle=0, \notag \\
&~&(L_{-1}^2-W_{-2}+\frac{16}{c}L_{-2} \mid \hat \psi \rangle=0, \notag \\
&~&(L_{-1}^3+\frac{24}{c}L_{-2}L_{-1}+\frac{12}{c}L_{-3}+\frac{24}{c}W_{-3} \mid \hat \psi \rangle=0. \eea
Inserting the null states into the correlation function, we get three differential equations on the correlation function involving the operator $\hat \psi$ corresponding to the state $|\hat \psi\rangle$. In particular the third equation can be transformed to
\be\label{diff1} \psi^{'''}(z,y)+\frac{24}{c}T(z,y)\psi^{'}(z,y)+\frac{12}{c}T^{'}(z,y)\psi(z,y)+\frac{24}{c}W(z,y)\psi(z,y)=0, \ee
where the prime denotes the derivative with respect to $z$ and
\be\label{psidef} \psi(z,y)=\frac{\langle e^{2\pi i \alpha{\cal{W}}_0-2\pi i\bar{\alpha}\bar{\cal{W}}_0} {\hat \psi}(z,y)...\rangle }
{\langle e^{2\pi i \alpha{\cal{W}}_0-2\pi i\bar{\alpha}\bar{\cal{W}}_0} ...\rangle}, \ee
and similarly
\bea
T(z,y)&=&\frac{\langle e^{2\pi i \alpha{\cal{W}}_0-2\pi i\bar{\alpha}\bar{\cal{W}}_0} \hat T(z,y)...\rangle }
{\langle e^{2\pi i \alpha{\cal{W}}_0-2\pi i\bar{\alpha}\bar{\cal{W}}_0} ...\rangle},\nn\\
W(z,y)&=&\frac{\langle e^{2\pi i \alpha{\cal{W}}_0-2\pi i\bar{\alpha}\bar{\cal{W}}_0}  \hat W_{-3}(z,y)...\rangle }
{\langle e^{2\pi i \alpha{\cal{W}}_0-2\pi i\bar{\alpha}\bar{\cal{W}}_0} ...\rangle}.
\eea
 The ellipsis  in (\ref{psidef}) denotes other local operators at $(z_i,y_i)$.  Unlike the Virasoro case without deformation, the functions $T(z,y)$ and $W(z,y)$ can not be determined simply by imposing the doubly periodic condition.

With the $sl(3,R)$ algebra the equation (\ref{diff1}) can be rewritten in a compact form
\be\label{diff3} \frac{\p \Psi(z,y)}{\p z}=a(z,y)\Psi(z,y), \ee
where
\be a(z,y)=(L_1+\frac{6}{c}T(z,y)L_{-1}-\frac{6}{c}W(z,y)W_{-2})
=\left( \begin{array}{ccc}
0&-\frac{12}{c}T(z,y)&-\frac{24}{c}W(z,y) \\
1&0&-\frac{12}{c}T(z,y) \\
0&1&0 \end{array} \right), \ee
\be \Psi(z,y)=
\left( \begin{array}{ccc}
\psi''(z,y)+\frac{2}{k}T(z,y)\psi(z,y) \\
\psi'(z,y)\\
\psi(z,y) \end{array} \right). \ee

From the definition of $\hat \psi(z,y)$,  we have
\be \frac{\partial \hat \psi(y,z)}{\partial y}=e^{-iy{\cal{W}}_0}[-i{\cal{W}}_0,\hat \psi(z)]e^{iy{\cal{W}}_0}
=e^{-iy{\cal{W}}_0}(-\hat W_{-2}\hat \psi(z))e^{iy{\cal{W}}_0}, \ee
where $(\hat W_{-2}\psi)(z)$ is the corresponding vertex operator for the state $\hat W_{-2}\mid \psi\rangle$.
Using the relation (\ref{nullvector}), we have
\be\label{diff2} \frac{\partial \Psi(z,y)}{\partial y}=b(z,y) \Psi(z,y), \ee
where
\be b=-\left( \begin{array}{ccc}
\frac{4}{c}T(z,y)&-\frac{4}{c}T'(z,y)-\frac{24}{c}W(z,y)&
\frac{4}{c}T''(z,y)+\frac{144}{c^2}T(z,y)^2 \\
0&-\frac{8}{c}T(z,y)&\frac{4}{c}T'(z,y)-\frac{24}{c}W(z,y) \\
1&0&\frac{4}{c}T(z,y) \end{array} \right). \ee

If the insertion of the operator $\hat \psi$ is away from the position of the non-local operator, the equations (\ref{diff3}) and (\ref{diff2}) can be solved formally by introducing an evolution operator,
 \be\label{solution} \Psi(z,y)=U(z,y;z_0,y_0)\Psi(z_0,y_0), \ee
where
 \be U(z,y;z_0,y_0)=P\exp[\int_{(z_0,y_0)}^{(z,y)} a(z,y)dz+b(z,y)dy] , \ee
 is a path-ordered integral on a contour in the two-dimensional complex plane. The consistency condition for the path-ordered integral is
\be\label{consistency} -\frac{\partial}{\partial y}a(z,y)+\frac{\partial}{\partial z}b(z,y)
-[a(z,y),b(z,y)]=0, \ee
or explicitly  as
\bea\label{consistency1} &~& \frac{\partial T}{\partial y}+2\frac{\partial W}{\partial z}=0, \notag \\
 &~&\frac{\partial^3 T}{\partial z^3}-6\frac{\partial W}{\partial y}
 +\frac{96}{c}T\frac{\partial T}{\partial z}=0,
 \eea
 which can be derived directly by ${\cal{W}}$ algebra in the large $c$ limit.
In Eq.  (\ref{solution}), $z$ and $y$ have to be regarded as two independent complex coordinates, representing the  evolution by ${\cal{L}}_0$ and ${\cal{W}}_0$ respectively.

Now let us discuss the monodromy condition. In the case at hand, there are two types of expansions:   one of them is the operator product expansion of two operators, and the other one is for inserting a complete bases along the thermal circle.  Correspondingly, we need to impose the monodromy condition on two circles,  the thermal circle and the circle  enclosing the two operators.  At each of the circle there is a monodromy condition. Because we only keep the vacuum module in the OPE of two operators, the monodromy around the the circle enclosing the two operators is trivial. The monodromy along the thermal cycle is more subtle.
By inserting a complete set of state basis the torus is cut open and becomes a cylinder, we can take a conformal transformation
\be w=e^{\frac{2\pi}{\beta}z},\ee
which maps the cylinder to the full complex plane. The operators are related by
\be \hat{\psi}(z)=(\frac{\partial w}{\partial z})^{2h} \hat{\psi}(w)
=(\frac{2\pi}{\beta})^{2h}e^{\frac{2\pi}{\beta}hz}\hat{\psi}(w).\ee
The monodromy in the $w$ coordinate is trivial. While in the $z$ coordinate, there is an extra phase $e^{\frac{2\pi}{\beta}hz}$ from the conformal transformation. When the conformal dimension of $\hat \psi$ is a half-integer,  the monodromy around the thermal circle is -1, as in \cite{Barrella:2013wja}.  When the conformal dimension of $\hat \psi$ is an integer, as in our case, the monodromy around the thermal circle must be trivial
\be \Psi(z+2\pi  \tau)_{\mbox{\tiny upper}}=\Psi(z)_{\mbox{\tiny lower}}.\ee
Considering the relation (\ref{evolvepath}), we have the monodromy condition around the thermal cycle
\be\label{monodromy} \Psi(z+2\pi\tau,y+2\pi \alpha)_{\mbox{\tiny lower}}=\Psi(z,y)_{\mbox{\tiny lower}}. \ee

\section{Monodromy analysis}

In this section, we use the monodromy condition to compute the correlation function (\ref{twopointT}) on the torus. Firstly we study the function in the denominator
$\langle e^{2\pi i\alpha{\cal{W}}_0-2\pi i \bar{\alpha}\bar{\cal{W}}_0}\rangle_{2\pi \tau}$, which is just the partition function of a higher spin black hole. We discuss the expectation values of the stress tensor and the higher spin charge. Then we compute  the two-point function in the numerator. In the discussion, we assume the  conformal dimension of the operator is of order $c$ but still light compared to the charge of the higher spin black hole so that we can ignore the backreaction to the background.

\subsection{Thermodynamics of the ensemble}

In this subsection, we show that  the monodromy condition can determine the thermodynamics of the ensemble with the higher spin deformation. The thermodynamics of higher spin black hole was studied holographically in \cite{Gutperle:2011kf} and in \cite{deBoer:2014fra} from the point of view of canonical deformation. A field theory derivation was presented in \cite{Kraus:2011ds} by using the perturbation expansion. Here we give another field theory derivation for the thermodynamics. In our derivation  the relation with the holographic study becomes more clear.

In this case, because of the translation invariance, the matrices $a$ and $b$ are constant-valued
\be a_0=
{\left( \begin{array}{ccc}
0&-\frac{12}{c}\langle T \rangle_0&-\frac{24}{c}\langle W \rangle_0 \\
1&0&-\frac{12}{c}\langle T \rangle_0 \\
0&1&0 \end{array} \right)}, \ee
\be b_0=
-{\left( \begin{array}{ccc}
\frac{4}{c}\langle T \rangle_0 &-\frac{24}{c}\langle W \rangle_0& \frac{144}{c^2}\langle T \rangle_0^2 \\
0& -\frac{8}{c}\langle T\rangle_0& -\frac{24}{c}\langle W \rangle_0 \\
1&0&\frac{4}{c}\langle T\rangle_0 \end{array} \right)}. \ee
Here the subscript $0$ denotes the expectation value with no operator insertion. Because the operators ${\cal{L}}$ and ${\cal{W}}$ commute, the matrices $a_0$ and $b_0$ also commute with each other. Then the monodromy condition (\ref{monodromy}) can be written as
\be \exp[2\pi \alpha b_0+2\pi \tau a_0]=1. \ee
This is exactly the monodromy condition suggested in \cite{Gutperle:2011kf}. Now we derive it from the field theory. With this monodromy condition we can easily solve the expectation value and derive the thermodynamics law as in \cite{Gutperle:2011kf}. Here we omit the details.

For our later study, we diagonalize the matrices $a_0$ and $b_0$ by
\be a_0=M
\left( \begin{array}{ccc}
\lambda_1&0&0 \\
0&\lambda_2&0 \\
0&0&\lambda_3 \end{array} \right) M^{-1}, \ee
\be b_0=-M
\left( \begin{array}{ccc}
\frac{1}{3}(\lambda_2+\lambda_3)^2+\frac{2}{3}\lambda_2\lambda_3&0&0 \\
0&\frac{1}{3}(\lambda_1+\lambda_3)^2+\frac{2}{3}\lambda_1\lambda_3&0 \\
0&0&\frac{1}{3}(\lambda_1+\lambda_2)^2+\frac{2}{3}\lambda_1\lambda_2 \end{array} \right)
M^{-1}, \ee
where
\be M=
\left( \begin{array}{ccc}
\frac{3}{4}\lambda_1^2-\frac{1}{4}\lambda_2^2-\frac{1}{4}\lambda_3^2&
\frac{3}{4}\lambda_2^2-\frac{1}{4}\lambda_1^2-\frac{1}{4}\lambda_3^2&
\frac{3}{4}\lambda_3^2-\frac{1}{4}\lambda_1^2-\frac{1}{4}\lambda_2^2\\
\lambda_1&\lambda_2&\lambda_3\\
1&1&1 \end{array} \right). \ee
Here $\lambda_1,\lambda_2,\lambda_3$ are three roots of the following cubic equation
\be \lambda^3+\frac{24}{c}\langle T\rangle_0\lambda+\frac{24}{c}\langle W\rangle_0=0, \ee
with $\lambda_1>\lambda_2>\lambda_3$.

\subsection{Two-point function}

In this subsection, we evaluate the correlation function (\ref{twopointT}) by imposing the monodromy condition. The conformal dimensions and the higher spin charges of two operators are respectively $(h_1,q_1)$ $(h_2,q_2)$. As we are interested in the entanglement entropy, we set
\bea h_1&=&h_2, \notag \\
q_1&=&-q_2, \eea
such that the operators can fuse to the vacuum module. We only consider light operators with $1<<h,q<<c$,  so that we can use the saddle point approximation and ignore their back reaction to the background. When we calculate the entanglement entropy, in $n\rightarrow 1$ limit, the twist operator satisfy the light operator condition.

Because $h$ and $q$ are much smaller than $a_0$ and $b_0$ which are the charges with no operator inserting, we can take a linear perturbation about the solution $a_0$, $b_0$ as
\bea a&=&a_0+a_1, \notag \\
 b&=&b_0+b_1, \eea
where
\be a_1=\left( \begin{array}{ccc}
0&-\frac{12}{c}(T-\langle T\rangle_0)& -\frac{24}{c}(W-\langle W\rangle_0) \\
0&0&-\frac{12}{c}(T-\langle T\rangle_0) \\
0&0&0 \end{array} \right),\nn \ee
\be b_1=- \left( \begin{array}{ccc}
\frac{4}{c}(T-\langle T\rangle_0)&-\frac{4}{c}T'-\frac{24}{c}(W-\langle W\rangle_0)&
\frac{4}{c}T''+\frac{288}{c^2}\langle T\rangle_0(T-\langle T\rangle_0) \\
0&-\frac{8}{c}(T-\langle T\rangle_0)&\frac{4}{c}T'-\frac{24}{c}(W-\langle W\rangle_0) \\
0&0&\frac{4}{c}(T-\langle T\rangle_0) \end{array} \right).\nn
\ee

Define
\bea U_0(z,y)&=&\exp[a_0 (z-z_0)+ b_0 (y-y_0)], \notag \\
U(z,y)&=&U_0(z,y)U_1(z,y). \eea
The differential equations (\ref{diff3}) and (\ref{diff2}) can be rewritten as
\bea &~&\frac{\partial}{\partial z} U_1(z,y)=U_0^{-1}(z,y) a_1(z,y) U_0(z,y), \notag \\
&~&\frac{\partial}{\partial y} U_1(z,y)=U_0^{-1}(z,y) b_1(z,y) U_0(z,y). \eea
The equations can be solved by
\be U_1(z,y)=P \exp \int_{\{z_0,y_0\}}^{\{z,y\}}
 [U_0^{-1}(z,y)a_1(z,y)U_0(z,y)dz+U_0^{-1}(z,y)b_1(z,y)U_0(z,y)dy]. \ee
Because of the consistency relation (\ref{consistency}) and (\ref{consistency1}) we can continuously deform the contour as long as the contour is away from the singular points.
The singular points of $a(z,y)$ and $b(z,y)$ can only appear at the locations of the operators, $z_1$ and $z_2$. By OPE we also have
\bea &~& T(z,y_1)\sim \frac{h_1}{(z-z_1)^2}+\frac{r_1}{z-z_1}, \notag \\
&~& T(z,y_2)\sim \frac{h_2}{(z-z_2)^2}+\frac{r_2}{z-z_2}, \notag \\
&~& W(z,y_1)\sim \frac{q_1}{(z-z_1)^3}+\frac{p_1}{(z-z_1)^2}+\frac{s_1}{z-z_1}, \notag \\
&~& W(z,y_2)\sim \frac{q_2}{(z-z_2)^3}+\frac{p_2}{(z-z_2)^2}+\frac{s_2}{z-z_2}. \eea
To the linear order the monodromy condition can be written as
\be\label{monodromy1} \oint U_0^{-1}(z,y)a_1(z,y)U_0(z,y)dz+U_0^{-1}(z,y)b_1(z,y)U_0(z,y)dy=0. \ee
This condition should satisfy for both the contour around the two operators and the contour around the thermal circle. Now we choose a special contour as follows
\be (z,y)=\left\{
\begin{aligned}
(z_2-\epsilon,y_2+t_1(y_1-y_2))~~~0<t_1<1 \\
(z_2-\epsilon+(z_1-z_2+2\epsilon)t_2,y_1)~~~0<t_2<1 \\
(z_1+\epsilon*e^{2\pi i*t_3},y_1)~~~0<t_3<1 \\
(z_1+\epsilon+(z_2-z_1-2\epsilon)t_4,y_1)~~~0<t_4<1 \\
(z_2-\epsilon,y_1+t_5(y_2-y_1))~~~0<t_5<1 \\
(z_2-\epsilon*e^{2\pi i t_6},y_2)~~~0<t_6<1 .
\end{aligned}
 \right.
\ee
The integrals from $t_1$, $t_5$ and $t_2$, $t_4$ are canceled with each other. The integrals from $t_3$ and $t_6$ can be evaluated by using the residue theorem.
Then the monodromy condition (\ref{monodromy1}) leads to
\be \sum_{i=1,2} Res_{z_i}(D(z,y_i)^{-1}M^{-1}a_0MD(z,y_i))=0, \ee
where
\be D(z,y)=\mbox{Diag}\left(e^{\lambda_1z-(\frac{1}{3}(\lambda_2+\lambda_3)^2+\frac{2}{3}\lambda_2\lambda_3)y},
e^{\lambda_2z-(\frac{1}{3}(\lambda_1+\lambda_3)^2+\frac{2}{3}\lambda_1\lambda_3)y},
e^{\lambda_3z-(\frac{1}{3}(\lambda_1+\lambda_2)^2+\frac{2}{3}\lambda_1\lambda_2)y}\right). \nn\ee
It is easy to solve these equations. The solution is
\be r_1=-r_2=m_1h_1+m_2q_1, ~~~s_1=-s_2=n_1h_1+n_2s_2, \ee
where
\bea
m_1&=&\frac{1}{2}(K_1-K_2)\nn\\
m_2&=&\frac{3}{2}(K_1+K_2)\nn\\
n_1&=&\frac{1}{2}(K_3-K_4)\nn\\
n_2&=&\frac{3}{2}(K_3+K_4)\nn
\eea
with
\bea K_1&=&
\frac{a_3(\lambda_1-\lambda_2)\lambda_3+a_1(\lambda_2-\lambda_3)\lambda_1+a_2(\lambda_3-\lambda_1)\lambda_2}
{a_3(\lambda_1-\lambda_2)+a_1(\lambda_2-\lambda_3)+a_2(\lambda_3-\lambda_1)} \notag \\
K_2&=&\frac{a_2a_3(\lambda_2-\lambda_3)\lambda_1+a_1a_2(\lambda_1-\lambda_2)\lambda_3
+a_3a_1(\lambda_3-\lambda_1)\lambda_2}
{a_2a_3(\lambda_2-\lambda_3)+a_1a_2(\lambda_1-\lambda_2)+a_3a_1(\lambda_3-\lambda_1)} \notag \\
 K_3&=&
\frac{a_3(\lambda_1-\lambda_2)(\frac{2}{3}\lambda_3^2-\frac{1}{3}\lambda_1^2-\frac{1}{3}\lambda_2^2)
+a_1(\lambda_2-\lambda_3)(\frac{2}{3}\lambda_1^2-\frac{1}{3}\lambda_2^2-\frac{1}{3}\lambda_3^2)
+a_2(\lambda_3-\lambda_1)(\frac{2}{3}\lambda_2^2-\frac{1}{3}\lambda_1^2-\frac{1}{3}\lambda_3^2)}
{a_3(\lambda_1-\lambda_2)+a_1(\lambda_2-\lambda_3)+a_2(\lambda_3-\lambda_1)} \notag \\
K_4&=&\frac{a_2a_3(\lambda_2-\lambda_3)(\frac{2\lambda_1^2}{3}-\frac{\lambda_2^2}{3}-\frac{\lambda_3^2}{3})
+a_1a_2(\lambda_1-\lambda_2)(\frac{2\lambda_3^2}{3}-\frac{\lambda_1^2}{3}-\frac{\lambda_2^2}{3})
+a_3a_1(\lambda_3-\lambda_1)(\frac{2\lambda_2^2}{3}-\frac{\lambda_1^2}{3}-\frac{\lambda_3^2}{3})}
{a_2a_3(\lambda_2-\lambda_3)+a_1a_2(\lambda_1-\lambda_2)+a_3a_1(\lambda_3-\lambda_1)} \notag \\
  a_1&=&\exp(-(z_1-z_2)\lambda_1
+(y_1-y_2)(\frac{2}{3}\lambda_1^2-\frac{1}{3}\lambda_2^2-\frac{1}{3}\lambda_3^2))\nn\\
a_2&=&\exp(-(z_1-z_2)\lambda_2+
(y_1-y_2)(\frac{2}{3}\lambda_2^2-\frac{1}{3}\lambda_1^2-\frac{1}{3}\lambda_3^2)) \nn\\
a_3&=&\exp(-(z_1-z_2)\lambda_3
+(y_1-y_2)(\frac{2}{3}\lambda_3^2-\frac{1}{3}\lambda_1^2-\frac{1}{3}\lambda_2^2)). \nn\eea
Using Eq. (\ref{evolveL} ) we find that the holomorphic part of the correlator (\ref{twopointT}) obeys the equation
\bea \frac{\partial}{\partial z_1} \log C_2(z_1,y_1;z_2,y_2) &=&r_1 \nn\\
\frac{\partial}{\partial y_1} \log C_2(z_1,y_1;z_2,y_2)&=&-s_1. \eea
Finally, we obtain
\bea
\lefteqn{\log C_2(z_1,y_1;z_2,y_2)} \notag \\
&=&-\frac{1}{2}\log[a_3(\lambda_1-\lambda_2)+a_1(\lambda_2-\lambda_3)+a_2(\lambda_3-\lambda_1)]
[a_1^{-1}(\lambda_2-\lambda_2)+a_2^{-1}(\lambda_3-\lambda_1)+a_3^{-1}(\lambda_1-\lambda_2)] h_1 \notag \\
&~&-\frac{3}{2}\log [a_3(\lambda_1-\lambda_2)+a_1(\lambda_2-\lambda_3)+a_2(\lambda_3-\lambda_1)]
[a_1^{-1}(\lambda_2-\lambda_2)+a_2^{-1}(\lambda_3-\lambda_1)+a_3^{-1}(\lambda_1-\lambda_2)]^{-1} q_1. \notag \\
\eea
We have similar result for the anti-holomorphic part.

\section{Holographic computation}
\subsection{Wilson line probe action}

The holographic computation of the correlation function (\ref{twopointT}) is to use the Wilson line proposal. The action of the Wilson line probe should give
the function $C_2$. The general framework for defining and computing the probe action can be found in \cite{Ammon:2013hba}.

To calculate the two-point function holographically we need the flat connection for the  spin-3 black hole
\be A=e^{-\rho L_0}(a+d)e^{\rho L_0} \ee
\be \bar{A}=e^{\rho L_0}(a+d)e^{-\rho L_0} \ee
with
\bea a&=&(L_1+\frac{6}{c}\langle T\rangle_0 L_{-1}-\frac{6}{c}\langle W\rangle_0 W_{-2})dz \notag \\
&~&-(W_2+\frac{12}{c}\langle T\rangle_0 W_0+\frac{36}{c^2}\langle T\rangle_0^2 W_{-2}
+\frac{12}{c}\langle W\rangle_0 L_{-1})dy \\
 \bar{a}&=&(L_{-1}+\frac{6}{c}\langle \tilde{T}\rangle_0 L_1-
\frac{6}{c}\langle \tilde{W}\rangle_0 W_2 )d\bar{z} \notag \\
&~&-(W_{-2}+\frac{12}{c}\langle \tilde{T}\rangle_0 W_0
+\frac{36}{c^2}\langle \tilde{T}\rangle_0^2W_2+\frac{12}{c}\langle \tilde{W}\rangle_0 L_1)d\bar{y}, \eea
 where $y=\frac{2\pi \a}{\b}t_E$. The terms proportional to $dy$ in $a, \bar{a}$ show that we are actually considering the black hole solution
with a chemical potential in the canonical boundary condition. Correspondingly, the dual CFT is canonically deformed by the spin-3 current. Here we use $y$ instead of $t$ just to show the relation with the field theory analysis more clearly.

To calculate the action of the Wilson line probe, we introduce
\be L=e^{-\rho L_0}e^{-(a_z z+a_y y)} ,\ee
 \be R=e^{\bar{a}_{\bar z}\bar{z}+\bar{a}_{\bar y}\bar{y}}e^{-\rho L_0} \ee
such that
\be A=LdL^{-1} ,\hs{3ex}
 \bar{A}=R^{-1}dR. \ee
 Then the probe action is defined by the diagonalized matrix
\bea H &\simeq& (L_iL_f^{-1}R_{f}^{-1}R_i)\nn\\
 &=&\mbox{diag}(t_1e^{4\rho_0},t_2,\frac{1}{t_1t_2}e^{-4\rho_0}), \eea
where the subscripts $i,f$ denote the endpoints of the Wilson line at the boundary; $t_1e^{4\r_0}, t_2, e^{-4\r_0}/t_1t_2$ are the eigenvalues of the matrix $L_iL_f^{-1}R_{f}^{-1}R_i$ and $\rho_0$ is the IR cut-off for the boundary. By direct calculation we get
\bea t_1&=&4[\frac{e^{\lambda_1(z_f-z_i)
+(-\frac{2}{3}\lambda_2\lambda_3-\frac{1}{3}(\lambda_2+\lambda_3)^2)(y_f-y_i)}}
{(\lambda_1-\lambda_2)(\lambda_1-\lambda_3)}
+\frac{e^{\lambda_2(z_f-z_i)
+(-\frac{2}{3}\lambda_1\lambda_3-\frac{1}{3}(\lambda_1+\lambda_3)^2)(y_f-y_i)}}
{(\lambda_2-\lambda_1)(\lambda_2-\lambda_3)} \notag \\
&~&+\frac{e^{\lambda_3(z_f-z_i)
+(-\frac{2}{3}\lambda_1\lambda_2-\frac{1}{3}(\lambda_1+\lambda_2)^2)(y_f-y_i)}}
{(\lambda_3-\lambda_1)(\lambda_3-\lambda_1)} ]\cdot
[\frac{e^{-\bar{\lambda}_1(\bar{z}_f-\bar{z}_i)+(\frac{2}{3}\bar{\lambda}_2\bar{\lambda}_3
+\frac{1}{3}(\bar{\lambda}_2+\bar{\lambda}_3)^2)(\bar{y}_f-\bar{y}_i)}}
{(\bar{\lambda}_1-\bar{\lambda}_2)(\bar{\lambda}_1-\bar{\lambda}_3)} \notag \\
&~&+\frac{e^{-\bar{\lambda}_2(\bar{z}_f-\bar{z}_i)+(\frac{2}{3}\bar{\lambda}_1\bar{\lambda}_3
+\frac{1}{3}(\bar{\lambda}_1+\bar{\lambda}_3)^2)(\bar{y}_f-\bar{y}_i)}}
{(\bar{\lambda}_2-\bar{\lambda}_1)(\bar{\lambda}_2-\bar{\lambda}_3)}
+\frac{e^{-\bar{\lambda}_3(\bar{z}_f-\bar{z}_i)+(\frac{2}{3}\bar{\lambda}_1\bar{\lambda}_2
+\frac{1}{3}(\bar{\lambda}_1+\bar{\lambda}_2)^2)(\bar{y}_f-\bar{y}_i)}}
{(\bar{\lambda}_3-\bar{\lambda}_1)(\bar{\lambda}_3-\bar{\lambda}_2)}], \nn
\eea

\bea t_1t_2&=&
4[\frac{e^{-\lambda_1(z_f-z_i)
+(\frac{2}{3}\lambda_2\lambda_3+\frac{1}{3}(\lambda_2+\lambda_3)^2)(y_f-y_i)}}
{(\lambda_1-\lambda_2)(\lambda_1-\lambda_3)}
+\frac{e^{-\lambda_2(z_f-z_i)
+(\frac{2}{3}\lambda_1\lambda_3+\frac{1}{3}(\lambda_1+\lambda_3)^2)(y_f-y_i)}}
{(\lambda_2-\lambda_1)(\lambda_2-\lambda_3)} \notag \\
&~&+\frac{e^{-\lambda_3(z_f-z_i)
+(\frac{2}{3}\lambda_1\lambda_2+\frac{1}{3}(\lambda_1+\lambda_2)^2)(y_f-y_i)}}
{(\lambda_3-\lambda_1)(\lambda_3-\lambda_1)} ]\cdot
[\frac{e^{\bar{\lambda}_1(\bar{z}_f-\bar{z}_i)-(\frac{2}{3}\bar{\lambda}_2\bar{\lambda}_3
+\frac{1}{3}(\bar{\lambda}_2+\bar{\lambda}_3)^2)(\bar{y}_f-\bar{y}_i)}}
{(\bar{\lambda}_1-\bar{\lambda}_2)(\bar{\lambda}_1-\bar{\lambda}_3)} \notag \\
&~&+\frac{e^{\bar{\lambda}_2(\bar{z}_f-\bar{z}_i)-(\frac{2}{3}\bar{\lambda}_1\bar{\lambda}_3
+\frac{1}{3}(\bar{\lambda}_1+\bar{\lambda}_3)^2)(\bar{y}_f-\bar{y}_i)}}
{(\bar{\lambda}_2-\bar{\lambda}_1)(\bar{\lambda}_2-\bar{\lambda}_3)}
+\frac{e^{\bar{\lambda}_3(\bar{z}_f-\bar{z}_i)-(\frac{2}{3}\bar{\lambda}_1\bar{\lambda}_2
+\frac{1}{3}(\bar{\lambda}_1+\bar{\lambda}_2)^2)(\bar{y}_f-\bar{y}_i)}}
{(\bar{\lambda}_3-\bar{\lambda}_1)(\bar{\lambda}_3-\bar{\lambda}_2)}]\nn
\eea
The action of the probe is given by \cite{deBoer:2014sna}
\be
I_{probe}=\Tr\left(\log(H)(\frac{h_1}{2}L_0+\frac{3q_1}{2}W_0)\right) .
\ee
It is straightforward to check that
\be C_2=e^{-I_{probe}}.  \ee

The agreement between the correlation function in the field theory and its holographic computation is remarkable.
The correlation function of two light operators is defined on the torus, and there is no restriction on the locations of the operators.
When one considers the entanglement entropy, the operators are set to the same time slice. In this case, the correlation function  is independent of the chemical potential and reduces to the one found in \cite{deBoer:2014sna}. In other words, the holographic computation of the entanglement entropy is not
sensitive to the choice of the gauge potential with or without the chemical potential. This is not the case if one considers more general two-point function on the torus.

\subsection{Holographic correspondence for picture transformation}

On the field side,  we can transform  two primary operators in the deformed theory into two descendant operators in a non-deformed theory by the picture-changing transformation. In this subsection, we would like to discuss the holographic correspondence of the picture-changing transformation. We suggest that the picture-changing transformation in the field theory correspond to a time-dependent gauge transformation on the gauge potential in the bulk.

Let us focus on a simple case. We assume that the state in the field theory is translational invariant. In a canonical deformed theory its holographic dual is just like the higher spin black hole as in \cite{Kraus:2011ds,deBoer:2014fra}.
\be A=e^{-\rho L_0}(a+d)e^{\rho L_0}, \ee
where
\bea\label{a} a&=&(L_1+\frac{6}{c}\langle T\rangle_0L_{-1}-\frac{6}{c}\langle W\rangle_0W_{-2} )dz \notag \\
&~&+\mu (W_2+\frac{12}{c}\langle T\rangle_0W_0+\frac{36}{c^2}W_{-2}+\frac{12}{c}\langle W\rangle_0L_{-1})dt. \eea
where $\langle T\rangle_0$ and $\langle W\rangle_0$ are constants. However, the state we consider here can be any translation invariant state, not necessarily the thermal state, so  $\langle T\rangle_0$ and  $ \langle W\rangle_0$ can take any values.

In the field theory we have $\cal{W}$ symmetry so that we can take a symmetry transformation on the state
\be\label{O} \mid \tilde{O}\rangle=e^{-i\lambda {\cal{W}}_0} \mid O\rangle. \ee
 In the gravity side, the transformation can be written as
\be\label{at} \tilde{a}=U^{-1}(a+d)U, \ee
where
\be U=\exp\left\{-\lambda(W_2+\frac{12}{c}\langle T\rangle_0W_0+\frac{36}{c^2}\langle T \rangle_0^2W_{-2}
+\frac{12}{c}\langle W\rangle_0L_{-1})\right\}, \ee
which is an asymptotic symmetry in the bulk.
This asymptotic symmetry was derived in \cite{Campoleoni:2010zq} for asymptotic AdS boundary condition and was extended to the canonical deformed boundary condition in \cite{Compere:2013gja}. In this simple case we can give the transformation explicitly. Furthermore, taking (\ref{a}) into (\ref{at}), we see that the gauge transformation keeps $a$ invariant. This corresponds to the fact that the state $|O \rangle$ is an eigenstate of the symmetry generator ${\cW}_0$.

In the above discussion, we  take the gauge transformation parameter $\l$ to be constant. We can furthermore extend it to be time dependent such that  it correspond to a picture-changing transformation in the field theory.\footnote{In \cite{deBoer:2014fra}, they regard the source term as a gauge field, and the time dependent transformation as the gauge transformation.} We take $\lambda=ts$, then the states transform as
\be \mid \tilde{O} \rangle=e^{-its{\cal{W}}_0} \mid O \rangle, \ee
and the corresponding operators transform as
\be \tilde{\phi}(t)=e^{-its{\cal{W}}_0}\phi(t)e^{its{\cal{W}}_0},\ee
which is exactly the picture-changing transformation.
Taking the parameter $\lambda=ts$ into (\ref{at}), we get
\be \tilde{a}=a-s(W_2+\frac{12}{c}\langle T\rangle_0W_0+\frac{36}{c^2}\langle T\rangle_0^2W_{-2}+\frac{12}{c}\langle W\rangle_0L_{-1})dt. \ee
For different $s$, it define a different picture in the field theory, and it corresponds to different asymptotic boundary condition. Specifically, when $s=\mu$, the gauge transformation cancel the chemical potential term in (\ref{a}),  and we get the gauge connection used in \cite{deBoer:2014sna}.

As we shown before, the probe action of the WL reproduces exactly the correlation function of two light operators located on the torus. One interesting question is if it is possible to read the correlation function from the gauge potential without chemical potential. The question is related to the holographic computation of two-point function of descendent operators. If these two operators are at the same slice, the direct computation of the probe action gives the correct answer. But if the operators are at different time slices, then one has to develop the WL proposal to address this issue.


\section{Conclusion and Discussion}

In this article, we studied the  entanglement entropy on a torus in the large $c$ CFT with $\cW$ current deformation.
More generally, we discussed the two-point function of the light operator with $h, q<<c$ under a thermal density matrix with a chemical potential.
Due to the presence of the deformation, the correlation function seems to be hard to compute. However, in the large $c$ limit, if we
accept that the vacuum module dominates the propagation, the problem is still tractable. First of all, under the limit, the ${\cW}_0$ charge commutes with the Hamiltonian so that we may apply a picture-changing transformation to turn off the deformation. Moreover, just as the Hamiltonian induce the translation of the time, the spin-3 operator induce the translation along an auxiliary coordinate. This leads us to find another differential equation on the wavefunction of the two-point function such that the monodromy problem could be well-defined. By imposing the monodromy condition, the two-point function could be determined at the linear order. Holographically we computed the probe action of the Wilson line in the background of spin-3 black hole with the chemical potential, and we found perfect agreement with field theory result.

Our treatment in the field theory could be applied to other deformation, as long as they are conserved.  In the large $c$ limit, the other higher spin currents  could be studied  straightforwardly.  This may help us to understand the $\cW$ conformal block on the torus\cite{Alkalaev:2016ptm}. Our study keeps to the leading order of $\frac{1}{c}$ expansion. It would be interesting to discuss the next leading order effect, especially considering the fact that the finite size correction to the entanglement entropy appears only at the next-leading order\cite{Cardy:2014jwa,Chen:2014unl}. On the bulk this corresponds to the 1-loop quantum correction to the holographic entanglement entropy.

Our study supports the Wilson line proposal to compute the holographic entanglement entropy. With canonical boundary condition, the probe action of the Wilson line computes the  HEE  even with non-zero chemical potential. This suggests that in the probe limit, the Wilson line proposal can be applied to more general bulk configuration\cite{Castro:2016ehj,Perlmutter:2016pkf} . On the other hand, how to determine the 1-loop quantum correction in the Wilson line proposal is an interesting question\cite{Besken:2016ooo}.


\vspace*{10mm}
\noindent {\large{\bf Acknowledgments}}\\

The work was in part supported by NSFC Grant No.~11275010, No.~11335012 and No.~11325522.
We would like to thank W. Song, A. Castro L. Hung and T. Takayanagi for helpful discussions. Wu would like to thank YITP for hospitality, where the final stage 
of this work was finished. Wu was supported by Short-term Overseas Research Program from Graduate School of Peking University.

\vspace*{5mm}

\begin{appendix}

\section{${\cal{W}}_3$ algebras}

For completeness we list the ${\cal{W}}_3$ algebras in this section as
\bea\label{Walgebra} T(z_1)T(z_2)&\sim& \frac{c/2}{(z_1-z_2)^4}+\frac{2T(z_2)}{(z_1-z_2)^2}
+\frac{\frac{\partial}{\partial z_2} T(z_2)}{(z_1-z_2)}+... , \notag \\
T(z_1)W(z_2)&\sim& \frac{3W(z_2)}{(z_1-z_2)^2}+\frac{\frac{\partial}{\partial z_2}W(z_2)}{(z_1-z_2)}+... , \notag \\
\frac{1}{N_3}W(z_1)W(z_2)&\sim& \frac{1}{(z_1-z_2)^6}+\frac{\frac{6}{c}T(z_2)}{(z_1-z_2)^4}+
\frac{\frac{3}{c}\frac{\partial}{\partial z_2}T(z_2)}{(z_1-z_2)^3}
+\frac{\frac{9}{10c}\frac{\partial^2}{\partial z_2^2}T(z_2)}{(z_1-z_2)^2}
+\frac{96}{c(5c+22)}\frac{\Lambda(z_2)}{(z_1-z_2)^2} \notag \\
&~&+\frac{1}{5c}\frac{\frac{\partial}{\partial z_2^3}T(z_2)}{z_1-z_2}
+\frac{48}{c(5c+22)}\frac{\frac{\partial}{\partial z_2}\Lambda (z_2)}{z_1-z_2}  +... ,
\eea
where $N_3=-\frac{5c}{6}$, and
\be \Lambda(z)=:T(z)^2:-\frac{3}{10}\partial^2T(z).  \ee
From the OPE coefficient, we can derive the commutators
\bea\label{Wcommutator} \left[L_m,L_n\right]&=&(m-n)L_{m+n}+\frac{c}{12}m(m^2-1)\delta_{m+n,0} , \notag \\
\left[ L_m,W_n \right]&=&(2m-n)W_{m+n} , \notag \\
\left[ W_m,W_n \right]&=&-\frac{1}{12}(m-n)(2m^2-mn+2n^2-8)L_{m+n}-\frac{40}{5c+22}(m-n)\Lambda_{m+n} \notag \\
&~&-\frac{5c}{6}\frac{1}{5!}m(m^2-1)(m^2-4) . \eea

\end{appendix}


\vspace*{5mm}
\begin{thebibliography}{99}

 \bibitem{Ryu:2006bv}
  S.~Ryu and T.~Takayanagi,
  ``Holographic derivation of entanglement entropy from AdS/CFT,''
  Phys.\ Rev.\ Lett.\  {\bf 96}, 181602 (2006)
  doi:10.1103/PhysRevLett.96.181602
  [hep-th/0603001].

  \bibitem{Lewkowycz:2013nqa}
  A.~Lewkowycz and J.~Maldacena,
  ``Generalized gravitational entropy,''
  JHEP {\bf 1308}, 090 (2013)
  doi:10.1007/JHEP08(2013)090
  [arXiv:1304.4926 [hep-th]].

  \bibitem{Faulkner:2013yia}
  T.~Faulkner,
  ``The Entanglement Renyi Entropies of Disjoint Intervals in AdS/CFT,''
  arXiv:1303.7221 [hep-th].

    \bibitem{Hartman:2013mia}
  T.~Hartman,
  ``Entanglement Entropy at Large Central Charge,''
  arXiv:1303.6955 [hep-th].

\bibitem{Krasnov:2000zq}
  K.~Krasnov,
  ``Holography and Riemann surfaces,''
  Adv.\ Theor.\ Math.\ Phys.\  {\bf 4}, 929 (2000)
  [hep-th/0005106].

  \bibitem{Headrick:2010zt}
  M.~Headrick,
  ``Entanglement Renyi entropies in holographic theories,''
  Phys.\ Rev.\ D {\bf 82}, 126010 (2010)
  [arXiv:1006.0047 [hep-th]].


  \bibitem{Barrella:2013wja}
  T.~Barrella, X.~Dong, S.~A.~Hartnoll and V.~L.~Martin,
  ``Holographic entanglement beyond classical gravity,''
  JHEP {\bf 1309}, 109 (2013)
  doi:10.1007/JHEP09(2013)109
  [arXiv:1306.4682 [hep-th]].

   \bibitem{Chen:2013kpa}
  B.~Chen and J.~J.~Zhang,
  ``On short interval expansion of R\'enyi entropy,''
  JHEP {\bf 1311}, 164 (2013)
  [arXiv:1309.5453 [hep-th]]. B.~Chen, F.~y.~Song and J.~j.~Zhang,
  ``Holographic Renyi entropy in AdS$_3$/LCFT$_2$ correspondence,''
  JHEP {\bf 1403}, 137 (2014)
  [arXiv:1401.0261 [hep-th]]. M.~Beccaria and G.~Macorini,
  ``On the next-to-leading holographic entanglement entropy in $AdS_{3}/CFT_{2}$,''
  JHEP {\bf 1404}, 045 (2014)
  [arXiv:1402.0659 [hep-th]]. Z.~Li and J.~j.~Zhang,
  ``On one-loop entanglement entropy of two short intervals from OPE of twist operators,''
  arXiv:1604.02779 [hep-th].

  \bibitem{Chen:2013dxa}
  B.~Chen, J.~Long and J.~j.~Zhang,
  ``Holographic R\'enyi entropy for CFT with W symmetry,''
  JHEP {\bf 1404}, 041 (2014)
  [arXiv:1312.5510 [hep-th]].

  \bibitem{Perlmutter:2013paa}
  E.~Perlmutter,
  ``Comments on Renyi entropy in AdS$_3$/CFT$_2$,''
  JHEP {\bf 1405}, 052 (2014)
  [arXiv:1312.5740 [hep-th]].


  \bibitem{Chen:2014unl}
  B.~Chen and J.~q.~Wu,
  ``Single interval Renyi entropy at low temperature,''
  JHEP {\bf 1408}, 032 (2014)
  [arXiv:1405.6254 [hep-th]].

  \bibitem{Chen:2015kua}
  B.~Chen and J.~q.~Wu,
  ``Holographic calculation for large interval R¨¦nyi entropy at high temperature,''
  Phys.\ Rev.\ D {\bf 92}, no. 10, 106001 (2015)
  doi:10.1103/PhysRevD.92.106001
  [arXiv:1506.03206 [hep-th]].


  \bibitem{Chen:2015uia}
  B.~Chen, J.~q.~Wu and Z.~c.~Zheng,
  ``Holographic R\'enyi Entropy of Single Interval on Torus: with W symmetry,''
  Phys.\ Rev.\ D {\bf 92}, 066002 (2015)
  [arXiv:1507.00183 [hep-th]].

  \bibitem{Zhang:2015hoa}
  J.~j.~Zhang,
  ``Holographic Rényi entropy for two-dimensional $ \mathcal{N}=\left(1,\;1\right) $ superconformal field theory,''
  JHEP {\bf 1512}, 027 (2015)
  doi:10.1007/JHEP12(2015)027
  [arXiv:1510.01423 [hep-th]].

  \bibitem{Yin:2007gv}
  X.~Yin,
  ``Partition Functions of Three-Dimensional Pure Gravity,''
  Commun.\ Num.\ Theor.\ Phys.\  {\bf 2}, 285 (2008)
  [arXiv:0710.2129 [hep-th]].

  \bibitem{Giombi:2008vd}
  S.~Giombi, A.~Maloney and X.~Yin,
  ``One-loop Partition Functions of 3D Gravity,''
  JHEP {\bf 0808}, 007 (2008)
  [arXiv:0804.1773 [hep-th]].


  \bibitem{Chen:2015uga}
  B.~Chen and J.~q.~Wu,
  ``1-loop partition function in AdS$_{3}$/CFT$_{2}$,''
  JHEP {\bf 1512}, 109 (2015)
  doi:10.1007/JHEP12(2015)109
  [arXiv:1509.02062 [hep-th]].

  \bibitem{Witten:1988hc}
  E.~Witten,
  ``(2+1)-Dimensional Gravity as an Exactly Soluble System,''
  Nucl.\ Phys.\ B {\bf 311}, 46 (1988).
  doi:10.1016/0550-3213(88)90143-5

   \bibitem{Campoleoni:2010zq}
  A.~Campoleoni, S.~Fredenhagen, S.~Pfenninger and S.~Theisen,
  ``Asymptotic symmetries of three-dimensional gravity coupled to higher-spin fields,''
  JHEP {\bf 1011}, 007 (2010)
  doi:10.1007/JHEP11(2010)007
  [arXiv:1008.4744 [hep-th]].

  \bibitem{Henneaux:2010xg}
  M.~Henneaux and S.~J.~Rey,
  ``Nonlinear $W_{infinity}$ as Asymptotic Symmetry of Three-Dimensional Higher Spin Anti-de Sitter Gravity,''
  JHEP {\bf 1012}, 007 (2010)
  doi:10.1007/JHEP12(2010)007
  [arXiv:1008.4579 [hep-th]].


\bibitem{Ammon:2013hba}
  M.~Ammon, A.~Castro and N.~Iqbal,
  ``Wilson Lines and Entanglement Entropy in Higher Spin Gravity,''
  JHEP {\bf 1310}, 110 (2013)
  doi:10.1007/JHEP10(2013)110
  [arXiv:1306.4338 [hep-th]].

  \bibitem{deBoer:2013vca}
  J.~de Boer and J.~I.~Jottar,
  ``Entanglement Entropy and Higher Spin Holography in AdS$_3$,''
  JHEP {\bf 1404}, 089 (2014)
  doi:10.1007/JHEP04(2014)089
  [arXiv:1306.4347 [hep-th]].

  \bibitem{deBoer:2014sna}
  J.~de Boer, A.~Castro, E.~Hijano, J.~I.~Jottar and P.~Kraus,
  ``Higher spin entanglement and $ {\mathcal{W}}_{\mathrm{N}} $ conformal blocks,''
  JHEP {\bf 1507}, 168 (2015)
  doi:10.1007/JHEP07(2015)168
  [arXiv:1412.7520 [hep-th]].


    \bibitem{deBoer:2014fra}
  J.~de Boer and J.~I.~Jottar,
  ``Boundary Conditions and Partition Functions in Higher Spin AdS$_3$/CFT$_2$,''
  arXiv:1407.3844 [hep-th].

\bibitem{Henneaux:2013dra} 
  M.~Henneaux, A.~Perez, D.~Tempo and R.~Troncoso,
  ``Chemical potentials in three-dimensional higher spin anti-de Sitter gravity,''
  JHEP {\bf 1312}, 048 (2013)
  doi:10.1007/JHEP12(2013)048
  [arXiv:1309.4362 [hep-th]].

  \bibitem{Long:2014oxa}
  J.~Long,
  ``Higher Spin Entanglement Entropy,''
  JHEP {\bf 1412}, 055 (2014)
  doi:10.1007/JHEP12(2014)055
  [arXiv:1408.1298 [hep-th]].

  \bibitem{Datta:2014ska}
  S.~Datta, J.~R.~David, M.~Ferlaino and S.~P.~Kumar,
  ``Higher spin entanglement entropy from CFT,''
  JHEP {\bf 1406}, 096 (2014)
  doi:10.1007/JHEP06(2014)096
  [arXiv:1402.0007 [hep-th]].

  \bibitem{Datta:2014uxa}
  S.~Datta, J.~R.~David, M.~Ferlaino and S.~P.~Kumar,
  ``Universal correction to higher spin entanglement entropy,''
  Phys.\ Rev.\ D {\bf 90}, no. 4, 041903 (2014)
  doi:10.1103/PhysRevD.90.041903
  [arXiv:1405.0015 [hep-th]].

  \bibitem{Fitzpatrick:2016ive}
  A.~L.~Fitzpatrick, J.~Kaplan, D.~Li and J.~Wang,
  ``On Information Loss in AdS$_3$/CFT$_2$,''
  arXiv:1603.08925 [hep-th].





    \bibitem{nielsen2010quantum}
M.~A. Nielsen and I.~L. Chuang, {\em Quantum computation and quantum
  information}.
\newblock Cambridge university press, 2010.


\bibitem{Holzhey:1994we}
  C.~Holzhey, F.~Larsen and F.~Wilczek,
  ``Geometric and renormalized entropy in conformal field theory,''
  Nucl.\ Phys.\ B {\bf 424}, 443 (1994)
  doi:10.1016/0550-3213(94)90402-2
  [hep-th/9403108].



    \bibitem{Hung:2014npa}
  L.~Y.~Hung, R.~C.~Myers and M.~Smolkin,
  ``Twist operators in higher dimensions,''
  JHEP {\bf 1410}, 178 (2014)
  doi:10.1007/JHEP10(2014)178
  [arXiv:1407.6429 [hep-th]].

   \bibitem{Harlow:2011ny}
  D.~Harlow, J.~Maltz and E.~Witten,
  ``Analytic Continuation of Liouville Theory,''
  JHEP {\bf 1112}, 071 (2011)
  doi:10.1007/JHEP12(2011)071
  [arXiv:1108.4417 [hep-th]].

  \bibitem{Eguchi:1986sb}
  T.~Eguchi and H.~Ooguri,
  ``Conformal and Current Algebras on General Riemann Surface,''
  Nucl.\ Phys.\ B {\bf 282}, 308 (1987).
  doi:10.1016/0550-3213(87)90686-9

\bibitem{Gutperle:2011kf}
  M.~Gutperle and P.~Kraus,
  ``Higher Spin Black Holes,''
  JHEP {\bf 1105}, 022 (2011)
  doi:10.1007/JHEP05(2011)022
  [arXiv:1103.4304 [hep-th]].

    \bibitem{Kraus:2011ds}
  P.~Kraus and E.~Perlmutter,
  ``Partition functions of higher spin black holes and their CFT duals,''
  JHEP {\bf 1111}, 061 (2011)
  doi:10.1007/JHEP11(2011)061
  [arXiv:1108.2567 [hep-th]].

  \bibitem{Alkalaev:2016ptm}
  K.~B.~Alkalaev and V.~A.~Belavin,
  ``Holographic interpretation of 1-point toroidal block in the semiclassical limit,''
  arXiv:1603.08440 [hep-th].

\bibitem{Cardy:2014jwa}
  J.~Cardy and C.~P.~Herzog,
  ``Universal Thermal Corrections to Single Interval Entanglement Entropy for Two Dimensional Conformal Field Theories,''
  Phys.\ Rev.\ Lett.\  {\bf 112}, no. 17, 171603 (2014)
  doi:10.1103/PhysRevLett.112.171603
  [arXiv:1403.0578 [hep-th]].

  \bibitem{Compere:2013gja}
  G.~Comp¨¨re and W.~Song,
  ``$\mathcal{W}$ symmetry and integrability of higher spin black holes,''
  JHEP {\bf 1309}, 144 (2013)
  doi:10.1007/JHEP09(2013)144
  [arXiv:1306.0014 [hep-th]].

   \bibitem{Castro:2016ehj}
  A.~Castro, N.~Iqbal and E.~Llabr¨¦s,
  ``Eternal Higher Spin Black Holes: a Thermofield Interpretation,''
  arXiv:1602.09057 [hep-th].
  
\bibitem{Perlmutter:2016pkf} 
  E.~Perlmutter,
  ``Bounding the Space of Holographic CFTs with Chaos,''
  arXiv:1602.08272 [hep-th].  

\bibitem{Besken:2016ooo}
  M.~Besken, A.~Hegde, E.~Hijano and P.~Kraus,
  ``Holographic conformal blocks from interacting Wilson lines,''
  arXiv:1603.07317 [hep-th].





\end{thebibliography}
\end{document}